\documentclass[a4paper,11pt]{article}
\pdfoutput=1 

\usepackage{jcappub} 
\usepackage{epstopdf}

\usepackage[T1]{fontenc} 

\title{\boldmath Non-thermal processes in standard big bang nucleosynthesis: III. Reactions with slow nuclei and the overall effect}

\author{V.T.Voronchev}
\affiliation{Skobeltsyn Institute of Nuclear Physics, Lomonosov Moscow State University,\\Leninskiye Gory 1, Moscow 119991, Russia}

\emailAdd{voronchev@srd.sinp.msu.ru}

\abstract{The present paper completes a series of our works on non-thermal nuclear processes in big bang nucleosynthesis (BBN) started in JCAP{\bf 05}(2008)010 (Part~I) and {\bf 05}(2009)001 (Part~II). The processes are triggered by non-Maxwellian particles naturally born in the main BBN reactions. Half of these reactions generate fast particles $k^+ (= n,p,t,\mathrm{^3He},\alpha )$. The other half, being radiative capture processes, produce slow nuclei $k^- (= d,t,\mathrm{^3He},\mathrm{^7Li},\mathrm{^7Be})$ which can undergo $(k^-,n)$ reactions with neutrons having large cross sections. The particle production rate $R_k$, thermalization time $\tau_k$, and effective number density $n_k$ are determined. It is shown that the values of $n_k$ at the Universe temperatures $T>65$~keV can exceed the number densities of Maxwellian $^7$Li and $^7$Be ions. To clarify the overall non-Maxwellian effect on BBN, both types of the non-Maxwellian particles are taken into account in the reaction network. Particular attention is paid to two-step sequential processes like $p(n,\gamma )d^-(n,\gamma )t$, $d(p,\gamma )\mathrm{^3He^-}(n,p)t$, $t(\alpha ,\gamma )\mathrm{^7Li^-}(n,\gamma )\mathrm{^8Li}$, $\mathrm{^3He}(\alpha ,\gamma )\mathrm{^7Be^-}(n,p)\mathrm{^7Li}$, $d(t,\alpha )n^+(A,n)a_1a_2$, and $d(\mathrm{^3He},\alpha )p^+(A,p)a_1a_2$ with $(A,a_1,a_2) = (\mathrm{^7Li},t,\alpha )$ and $(\mathrm{^7Be},\mathrm{^3He},\alpha )$. It is obtained that the non-Maxwellian particles can selectively affect the element abundances, e.g., improve the prediction on $^7$Li/H by $\sim 1.5\%$ and at the same time leave unchanged the $^4$He abundance. The main conclusion however is that these particles are unable to significantly change the standard picture of BBN in general, and provide a pathway toward a solution of the cosmological lithium problem in particular.}

\begin{document}
\maketitle
\flushbottom

\section{Introduction}
\label{sec:intro}

A key issue in the modeling of big bang nucleosynthesis (BBN) is a proper description of nuclear reaction kinetics in the primordial plasma. The rate equations for abundances $Y_i$ of elements $i$ involving in forward and reverse processes $N_i(i)+N_j(j) \rightleftarrows N_k(k)+N_l(l)$ are given by \cite{wago67,wago69}
\begin{equation}\label{eq:rate}
    \frac{Y_i}{dt} =
    \sum_{j,k,l} N_i
      \left (
          - \frac{Y^{N_i}_i Y^{N_j}_j}{N_i!N_j!} [ij]_k
          + \frac{Y^{N_l}_l Y^{N_k}_k}{N_l!N_k!} [lk]_j
      \right ),
\end{equation}
where $N_m$ is the number of nuclei $m$. For a two-body $i+j$ reaction, $[ij]_k = \rho_b N_A \langle\sigma v\rangle$ where the $i+j$ reactivity $\langle \sigma v\rangle$ is defined as $\sigma\times |\mathbf{v}_i-\mathbf{v}_j|$ folded over the reactant velocity distribution functions $f(\mathbf{v}_i)$ and $f(\mathbf{v}_j)$. This quantity is the major nuclear input in BBN simulations.

Equation \ref{eq:rate} is universal and can be applied to various scenarios of BBN. However, in different models the values of $[ij]_k$ may differ, depending on the adopted particle distribution functions. In the standard model of BBN, the particle distributions are assumed to be Maxwellian.

The phenomenon of non-Maxwellian particles in BBN is a widely discussed topic in the literature. They can appear, e.g., in scenarios beyond the Standard Model due to the decay of exotic relic particles \cite{cybu03,jeda04,kawa05,jeda06}. An acceleration of nucleons in the matter \cite{kang19} and a transient phenomenon \cite{park20} were also considered to be the sources of nonthermal particles. Models developed in \cite{bert13,hou17,kusa19} modify standard BBN within a thermal approach: although they do not involve any nonthermal particles, it is assumed that the primordial plasma species obey not Maxwellian but non-extensive Tsallis statistics. A combined Tsallis-Maxwellian approach was proposed in \cite{jang21}. Here, the photon distribution was parameterised with Tsallis statistics while particles still remain Maxwellian. A mechanism of non-Maxwellian distortion for particle distributions in the presence of a stochastic magnetic field was analyzed in \cite{luo19}. Useful links to other non-standard models of BBN can be found in a recent review \cite{pdg}.

Strictly speaking, even in standard BBN a small fraction of non-Maxwellian particles should also present in the primordial plasma as the products of nuclear reaction are not initially Maxwellian. BBN reactions have $Q$-values in the range of 1--20~MeV, so they generate energetic particles which can somewhat increase the population of the respective high-energy distribution tails. This was clearly demonstrated on the example of plasma neutrons \cite{boyd10,naka11}. In particular, it was shown that the non-Maxwellian neutron distribution tails created by fast $d+t$ and $d+d$ neutrons increase the rates of some reactions in the plasma. In addition to neutrons, it was found \cite{voro10,voro12}  that fast charged products of BBN reactions can also enhance nuclear processes.

All these studies discuss non-thermal effects triggered by MeV particles. At the same however half of the main BBN reactions are radiative capture processes $i(j,\gamma )k$ producing slow particles $k$ as almost all energy released is carried out by photons $\gamma$. These slow products could potentially enhance the low-energy tail of the $k$-particle distribution compared to a Maxwellian function. The enhancement may in principle affect the rates of $k+n$ reactions with neutrons as their cross sections, usually depending on energy as $\sigma \varpropto 1/\sqrt{E}$, are large at low energies. Unlike fast particle-induced reactions, little is still known about a possible role of these processes in BBN. The only study in the field was published recently \cite{voro24}. The situation becomes even more interesting if we remember that deuterons as well as $^7$Be are essentially synthesized in radiative capture reactions, so all these nuclei are initially slow in the primordial plasma.

The purpose of present paper is to incorporate sequential processes involving slow nuclei like $i(j,\gamma )k[\text{slow}]+n$ in a BBN reaction network and to draw an overall conclusion on the contribution of non-Maxwellian reaction products to BBN, completing a serious of our studies started in \cite{jcap08,jcap09}.

\begin{table}[t]
\centering
\begin{tabular}{|c|c|c|c|}
\hline
No. & Reaction & Product & Reactant \\
\hline
 & Top 11 reactions & & \\
1 & $p + n \rightleftarrows \gamma + d$ & s & f \\
2 & $d + p \rightleftarrows \gamma + \mathrm{^3He}$ & s & f \\
3 & $d +d \rightleftarrows n + \mathrm{^3He}$ & f & \\
4 & $d + d \rightleftarrows p + t$ & f & \\
5 & $t + d \rightleftarrows n + \alpha$ & f & f \\
6 & $\mathrm{^3He} + n \rightleftarrows p + t$ & f & f, s \\
7 & $\mathrm{^3He} + d \rightleftarrows p + \alpha$ & f & f \\
8 & $t + \alpha \rightleftarrows \gamma + \mathrm{^7Li}$ & s & f \\
9 & $\mathrm{^3He} + \alpha \rightleftarrows \gamma + \mathrm{^7Be}$ & s & f \\
10 & $\mathrm{^7Li} + p \rightleftarrows \alpha + \alpha$ & f & f \\
11 & $\mathrm{^7Be} + n \rightleftarrows p + \mathrm{^7Li}$ & f & f, s \\
 & Reactions with neutrons & & \\
12 & $d + n \rightleftarrows \gamma + t$ & & f, s \\
13 & $\mathrm{^3He} + n \rightleftarrows \gamma + \alpha$ & & f, s \\
14 & $\mathrm{^7Li} + n \rightleftarrows \gamma + \mathrm{^8Li}$ & & f, s \\
15 & $\mathrm{^7Be} + n \rightleftarrows \alpha + \alpha$ & & f, s\\
 & Other processes & & \\
16 & $t + p \rightleftarrows \gamma + \alpha$ & & f \\
17 & $t + t \rightarrow n +n + \alpha$ & & f \\
18 & $\mathrm{^3He} + \mathrm{He} \rightarrow p + p + \alpha$ & & f \\
19 & $\mathrm{^7Be} + p \rightleftarrows \gamma + \mathrm{^8B}$ & & f \\
20 & $\mathrm{^7Li} + \alpha \rightleftarrows \gamma + \mathrm{^{11}B}$ & & f \\
21 & $\mathrm{^7Be} + \alpha \rightleftarrows \gamma + \mathrm{^{11}C}$ & & f \\
22 & $d + n \rightarrow n + n + p$ & & f \\
23 & $d + p \rightarrow p + n + p$ & & f \\
24 & $\mathrm{^7Li} + n \rightarrow n + t + \alpha$ & & f \\
25 & $\mathrm{^7Li} + p \rightarrow p + t + \alpha$ & & f \\
26 & $\mathrm{^7Be} + n \rightarrow n + \mathrm{^3He} + \alpha$ & & f \\
27 & $\mathrm{^7Be} + p \rightarrow p + \mathrm{^3He} + \alpha$ & & f \\
\hline
\end{tabular}
\caption{\label{tab:list} BBN reactions for particles with mass number $A \leq 7$. Processes involving rare $^6$Li and low-probability three-body reactions are not listed. The third column identifies slow (s) and fast (f) products of the key 11 reactions controlling the primordial element production. The fourth column shows what types of the non-Maxwellian particles are taken into account in the processes of the second column.}
\end{table}
\begin{figure}
\begin{center}
\includegraphics{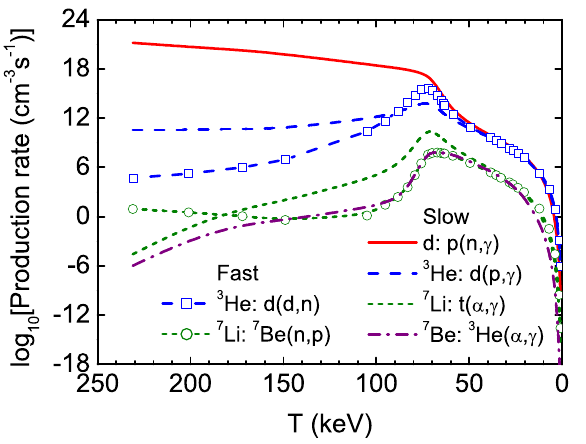}
\caption{\label{fig:Rk} The production rate of slow and fast $d$, $^3$He, $^7$Li, and $^7$Be as a function of the Universe temperature.}
\end{center}
\end{figure}

\section{Non-Maxwellian reaction products in the primordial plasma}
\label{sec:particles}

The list of BBN reactions that we will pay special attention to is presented in table~\ref{tab:list}. The third and fourth columns identify slow (s) and fast (f) products of the top 11 processes controlling the production of light elements. All these non-Maxwellian particles will be taken into account in the BBN reaction network.

\subsection{Slow nuclei}
\label{sec:slow}

The main source of slow nuclei in the primordial plasma is radiative capture reactions $i(j,\gamma )k$. In these processes, most of the energy released $E_i+E_j+Q$ is carried out by photons. With a precision of a few keV the photon energy is \cite{iliadis}
\begin{equation}\label{eq:Eg}
    E_\gamma = Q + E' + (Q+E')\frac{v_k}{c}\cos\theta
    - \frac{(Q+E')^{2}}{2m_k c^{2}}
\end{equation}
while the slow nucleus $k$ is born at the energy
\begin{equation}\label{eq:Ek}
    E_k =
    E_\text{c.m.} + E' + Q - E_\gamma .
\end{equation}
In these equations, $E_\text{c.m.}$ and $E'$ are the energy of the center-off-mass frame (c.m.) and the $(i+j)$ kinetic energy in the c.m. frame, respectively, $v_k$ is the velocity of the particle $k$ in the laboratory system, and $\theta$ is the direction of motion of the c.m. frame relative to the observer.

\begin{figure}
\begin{center}
\includegraphics{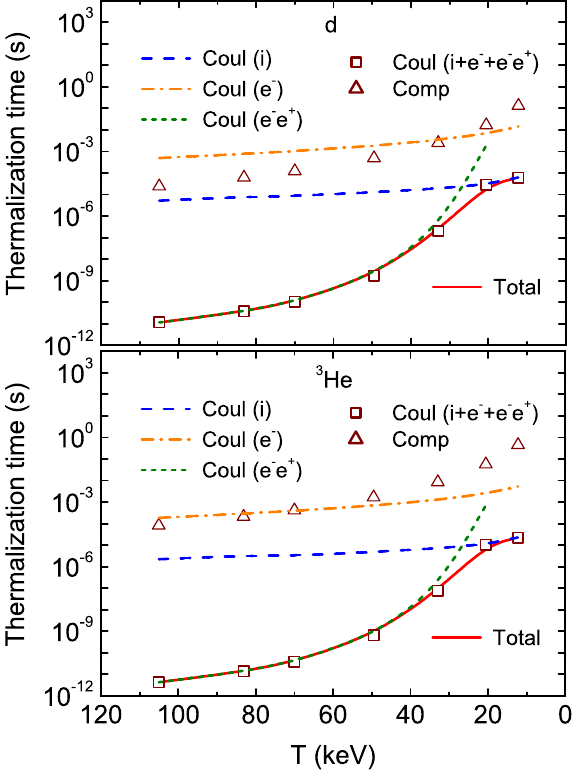}
\caption{\label{fig:tau23} The thermalization time of the deuteron and the $^3$He nucleus in the primordial plasma. The total and partial $\tau_k$ provided by different mechanisms are shown.}
\end{center}
\end{figure}

Almost half of the top BBN reactions produce slow nuclei $k = d$, $^3$He, $^7$Li, and $^7$Be (see table~\ref{tab:list}). Their production rate is
\begin{equation}\label{eq:Rk}
    R_{k,\text{slow}} =
    n_i n_j \langle\sigma v\rangle_{ij \rightarrow \gamma k},
\end{equation}
where $n_i$ and $n_j$ are the reactant number densities. In the present study, $n_i$ and $n_j$ were determined by running the Kawano's code \cite{kawano} with the baryon-to-photon ratio $\eta = n_\text{B}/n_\gamma = 6.09\times 10^{-10}$ and the values of $\langle\sigma v\rangle N_A$ for the main reactions taken from \cite{cybu04,ando06}. The production rate of slow $d$, $^3$He, $^7$Li, and $^7$Be as a function of the Universe temperature is plotted in figure \ref{fig:Rk}. For comparison, the rates $R_{k,\text{fast}}$ of competing nuclear channels generating the same nuclei at high energies are also shown. It is worth noting here that \emph{all} deuterons and \emph{all} $^7$Be are initially slow in the plasma while $^3$He and $^7$Li are predominantly produced through the `high-energy' channels. These nuclei may undergo reactions with neutrons (see table \ref{tab:list}) before their Maxwellization occurs. It is important to note that the cross sections of these reactions become large at low energies \cite{nndc}.

The abundance of slow $d$, $^3$He, $^7$Li, and $^7$Be depends on the time of their thermalization in the plasma. This process is governed by Coulomb interaction with background electrons (Coul($e^{-}$)), electron-positron pairs (Coul($e{^\pm}$)), ions (Coul($i$)), and by Compton scattering off thermal photons (Comp). These mechanisms are described, e.g., in \cite{kawa05,sivu66,reno88}. Besides, the process of nuclear elastic scattering off ambient nucleons and nuclei may also take place. However, this mechanism can be neglected as the particles are slow.

\begin{figure}
\begin{center}
\includegraphics{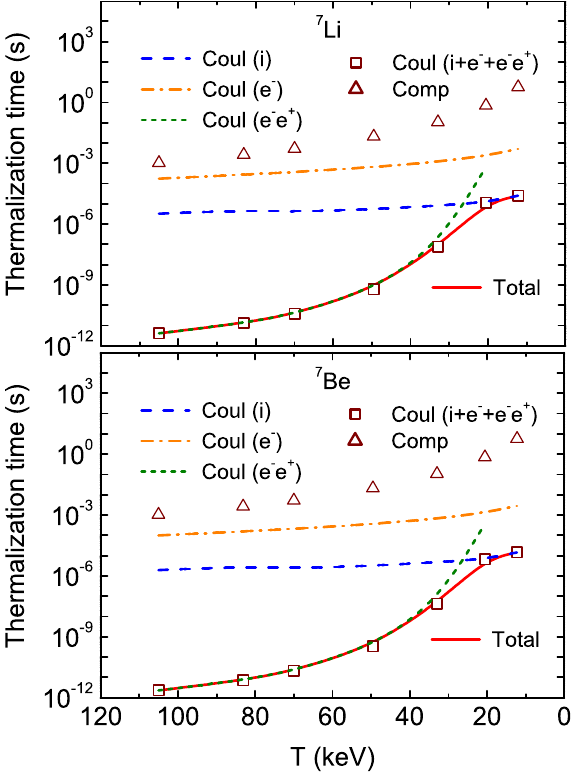}
\caption{\label{fig:tau7} The same parameters as in figure \ref{fig:tau23} for the $^7$Li and $^7$Be nuclei.}
\end{center}
\end{figure}

The calculated thermalization times $\tau_k$ are presented in figure~\ref{fig:tau23} (for deuterons and $^3$He) and figure \ref{fig:tau7} (for $^7$Li and $^7$Be). The total time together with partial times $\tau_k$ corresponding to the different interaction mechanisms are shown. It is seen that the most effective thermalization occurs due to scattering off electron-positron pairs at high temperatures $T \gtrsim 30$~keV and also due to collisions with ions at lower temperatures, after the $e^\pm$ annihilation is completed.

\begin{figure}
\begin{center}
\includegraphics{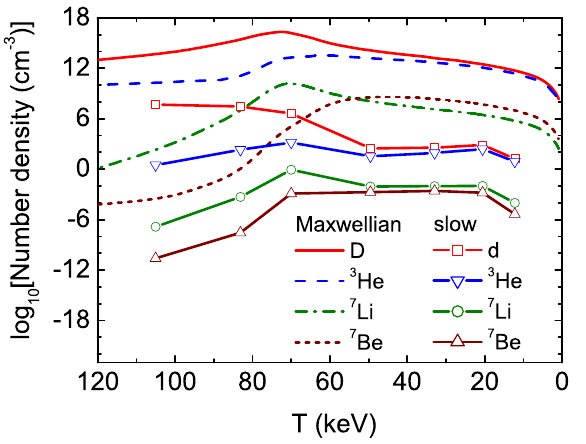}
\caption{\label{fig:den-slow} The number densities of slow and Maxwellian particles in the primordial plasma.}
\end{center}
\end{figure}

The effective number densities of the slow particles defined as $n_{k,\text{slow}} = R_{k,\text{slow}}\times\tau_k$  are given in figure~\ref{fig:den-slow} together with the respective number densities for Maxwellian nuclei. It is interesting that at an early stage of BBN the slow particle abundances prove to be larger that the abundances of Maxwellian $^7$Li and $^7$Be.

\subsection{Fast nucleons and nuclei}
\label{sec:fast}

\begin{figure}
\begin{center}
\includegraphics{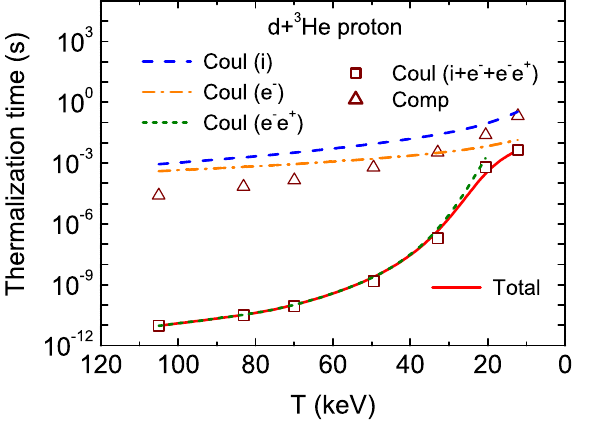}
\caption{\label{fig:tau1} The slowing-down time of the 14.7-MeV proton.}
\end{center}
\end{figure}
\begin{figure}
\begin{center}
\includegraphics{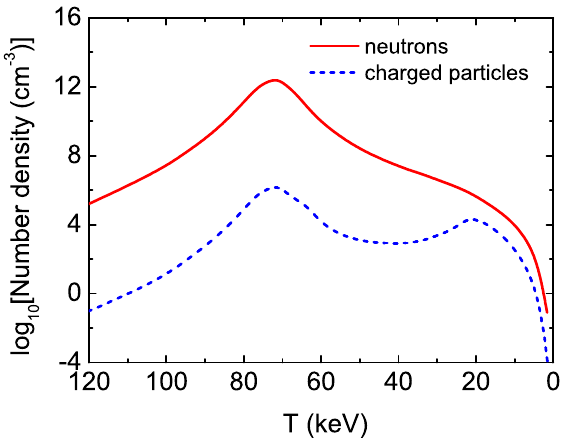}
\caption{\label{fig:den-fast} The number density of fast neutrons and fast charged particles.}
\end{center}
\end{figure}

To get an overall picture of the influence of non-Maxwellian reaction products on BBN, processes with fast neutrons and charged particles, such as protons, tritons, helions, and $\alpha$-particles (see table~\ref{tab:list}), were also taken into account.

The slowing-down process of fast ions in the primordial plasma is mainly governed by Coulomb scattering off electrons and electron-positron pairs. As an example, the total and partial thermalization times for 14.7-MeV $d+\mathrm{^3He}$ protons are plotted in figure~\ref{fig:tau1}. For fast neutrons (they are generated in the $d+t$ and $d+d$ reactions), the main slowing-down mechanisms are nuclear elastic scattering off thermal protons and magnetic-moment interaction with electrons and positrons \cite{jcap09}.

The number densities of fast neutrons and fast charged particles are given in figure~\ref{fig:den-fast}. A comparison of figure~\ref{fig:den-slow} and figure~\ref{fig:den-fast} indicates that the most abundant non-Maxwellian particles in the plasma are neutrons.

\section{Element abundances}
\label{sec:abun}

In the present BBN simulations, all forward and reverse reactions in table~\ref{tab:list} induced by both slow and fast particles were taken into account. Special attention was paid to the following sequential processes:

\emph{a}) for the slow particles, these are the production of slow nuclei in $i(j,\gamma )k$ followed by their burn-up in $n+k$ reactions with large cross sections
\begin{equation}\label{eq:seq1}
    p(n,\gamma )d\mathrm{[slow]}(n,\gamma )t,
\end{equation}
\begin{equation}\label{eq:seq2}
    d(p,\gamma )\mathrm{^3He[slow]}(n,p)t,
\end{equation}
\begin{equation}\label{eq:seq3}
    t(\alpha ,\gamma )\mathrm{^7Li[slow]}(n,\gamma )\mathrm{^8Li},
\end{equation}
\begin{equation}\label{eq:seq4}
    \mathrm{^3He}(\alpha ,\gamma )\mathrm{^7Be[slow]}(n,p)\mathrm{^7Li}
\end{equation}

\emph{b}) for the fast particles, these are the disintegration of loosely bound nuclei $A = d$, $^7$Li, and $^7$Be induced by fast neutrons and protons generated in the $d+t$ and $d+\mathrm{^3He}$ reactions
\begin{equation}\label{eq:seq5}
    d(t,\alpha )n\mathrm{[fast]}(A,n)a_1a_2,
\end{equation}
\begin{equation}\label{eq:seq6}
    d(\mathrm{^3He},\alpha )p\mathrm{[fast]}(A,p)a_1a_2,
\end{equation}
where $(A,a_1,a_2) = (\mathrm{^7Li},t,\alpha )$ and $(\mathrm{^7Be},\mathrm{^3He},\alpha )$. The $n+A$ rates were calculated using the reactivity $\langle\sigma v\rangle_{nA\rightarrow na_1a_2}$ obtained with the realistic neutron distribution functions having enhanced non-Maxwellian tails at MeV energies \cite{naka11}.

We note that some inaccuracy in the rates of processes \ref{eq:seq1}--\ref{eq:seq4} may come from choosing the initial energy $E_{k,0}$ of the particle $k$ as it depends on the $(i+j)$ center-off-mass velocity $\mathbf{V}_\text{c.m.} = m_i\mathbf{v}_i/(m_i+m_j) + m_j\mathbf{v}_j/(m_i+m_j)$. To clarify the point, several values of $V_\text{c.m.}$ corresponding to typical speeds of the reactants $m (= i,j)$ were tested. These are the most probable speed $v'_m$ following from the condition $df_i(v_m)/dv_m = 0$, the mean speed $\langle v_m\rangle$, the r.m.s. speed $\langle v^2_m\rangle^{1/2}$, and also the average relative velocity $\langle |\mathbf{v_i}-\mathbf{v_j}|\rangle$. It was obtained that the respective $n+k$ rates differ by no more than 5\%.

\begin{table}[t]
\centering
\begin{tabular}{|c|c|c|c|}
\hline
Element & \multicolumn{2}{c}{BBN scenario} |& Observations \\
 & with non-Maxwellian & Maxwellian & \\
\hline
D/H ($\times 10^{-5}$) & 2.576 & 2.575 & $2.547 \pm 0.029$ \\
$^{3}$He/H ($\times 10^{-5}$) & 1.009 & 1.011 & $<1.1 \pm 0.2$ \\
$Y_p$ ($\times 10^{-1}$) & 2.456 & 2.456 & $2.45 \pm 0.03$ \\
$^{7}$Li/H ($\times 10^{-10}$) & 4.365 & 4.424 & $1.6 \pm 0.3$ \\
$[\mathrm{^6Li} + A>7]$/H ($\times 10^{-14}$) & 1.237 & 1.236 & \\
\hline
\end{tabular}
\caption{\label{tab:abun} The element abundances in BBN with the non-Maxwellian particles in the primordial plasma. The Maxwellian scenario is also given for comparison. Observational data are taken from \cite{pdg}.}
\end{table}

\begin{figure}
\begin{center}
\includegraphics{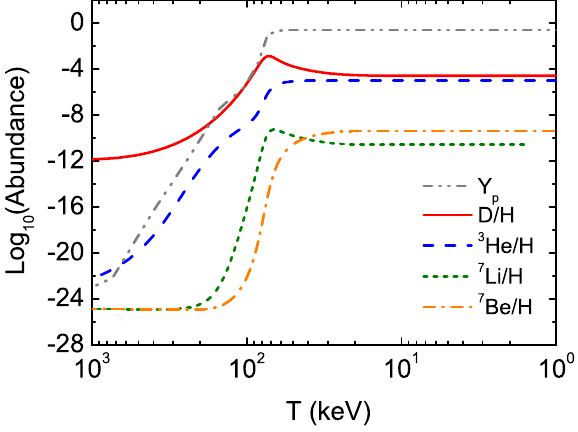}
\caption{\label{fig:Y} The dynamics of light element production as the Universe expands and cools.}
\end{center}
\end{figure}

\begin{figure}
\begin{center}
\includegraphics{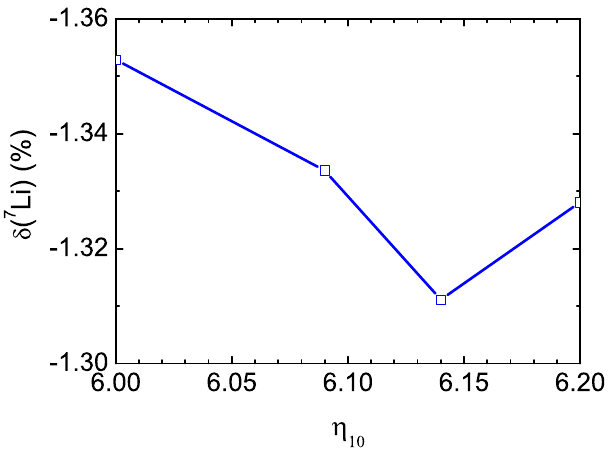}
\caption{\label{fig:ratio-7Li} The non-Maxwellian effect on the $^7$Li abundance at different values $\eta_{10} \equiv \eta\times 10^{10}$ (for $\delta$, see the text).}
\end{center}
\end{figure}

The element abundances A/H calculated with the baryon-to-photon ratio $\eta = 6.09\times 10^{-10}$ are given in table~\ref{tab:abun}. The scenarios allowing for and neglecting the non-Maxwellian particles are shown. A rather slight impact on the abundances is observed. In particular, the non-Maxwellian particles can improve the theoretical prediction on $^7$Li/H by $\sim 1.5\%$ and the same time leave unchanged the $^4$He abundance. These changes cannot be seen in figure~\ref{fig:Y} showing the element production dynamics as the Universe expands and cools. It was also obtained that the effect is practically independent of the baryon-to-photon ratio in the range of $6\times{10^{10}}$--$6.2\times{10^{10}}$. This is demonstrated in figure~\ref{fig:ratio-7Li} showing the non-Maxwellian $Y'$ to Maxwellian $Y$ abundance ratio $\delta(^7\mathrm{Li})=Y'/Y-1$ at different $\eta$.


\section{Conclusion}
\label{sec:conc}

The present paper completes the series of our works on non-thermal nuclear processes in the primordial plasma triggered by non-Maxwellian particles produced in the BBN reactions. The purpose was to determine the main characteristics of slow reaction products and to find whether they, together with fast particles, can affect the primordial element production. The results can be summarized as follows:
\begin{itemize}
\item[]1. It is shown that some amount of non-Maxwellian (both slow and fast) reaction products is continuously present in the plasma. Their number densities at the Universe temperature $T \gtrsim 50$~keV exceed the number densities of Maxwellian $^7$Li and $^7$Be and can be comparable to the Maxwellian $^3$He number density at $T \simeq 70$--80~keV. The most abundant non-Maxwellian particles are neutrons. \\
\item[]2. It is found that these particles can selectively, but only slightly, alter the element abundances compared to the Maxwellian values. For example, they can improve the theoretical prediction on $^7$Li, reducing its abundance by $\sim 1.5\%$. The abundance of $^4$He remains unchanged. \\
\end{itemize}

The overall conclusion however is that neither slow nor fast reaction products are able to significantly change the standard picture of BBN in general, and provide a pathway toward a solution of the cosmological lithium problem in particular.

\end{document}